\documentclass[journal,draftclsnofoot,onecolumn,12pt]{IEEEtran}

\usepackage[update,prepend]{epstopdf}
\usepackage{array,arydshln}
\usepackage{graphics}
\usepackage{multirow}
\usepackage{tikz}
\usepackage{bbm} 
\usepackage{pdfpages}
\usepackage{multirow}
\usepackage{subfig}
\usepackage{comment}

\usepackage{setspace}	
\usepackage{graphicx}
\usepackage{algorithm,algorithmic}
\usepackage{multicol}

\usepackage[justification=centering]{caption}
\usepackage{textcomp}
\usepackage{psfrag}
\usepackage{arydshln}
\usepackage{url}
\usepackage{soul}
\usepackage{graphicx,color}
\usepackage[nolist]{acronym}

\usepackage{mathtools,lipsum}
\usepackage{cuted}
\usepackage{amsmath}
\usepackage{graphicx}
\usepackage{booktabs}
\usepackage{float}

\begin{document}
	\graphicspath{{./Figures/}}
	
	\title{
	SG-Based Analysis of LEO Satellite-Relayed Communication Systems
	}
	
	\author{Yanbo Cao, Derui Gao and Jingyu Xu
	\thanks{Every author has the equal contribution in this work. The authors are with University of Electronic Science \& Technology of China, Qingshuihe Campus, No. 2006, Xiyuan Ave., West High-tech Zone, Chengdu, Sichuan, China. (e-mail:caoyanbo2000@foxmail.com; 2562189344@qq.com; jingyuxu.claire@foxmail.com)
	\vspace{-16mm}
}
}	
	
	\maketitle
	\begin{abstract}
		Due to their low latency, high capacity, and seamless worldwide coverage, low Earth orbit (LEO) satellites are essential to the equal access network. Stochastic geometry (SG) is an appropriate method for such a large and irregular system. The SG model can effectively assess and estimate the performance of the network as well as handle the growing network scale. In this article, a number of common satellite distribution models are examined. In the non-technical description, system-level metrics such as coverage probability are introduced. The impact of gateway density, as well as the quantity and height of satellites, on latency and likelihood of coverage, is then researched. This essay concludes by outlining potential uses for SG in the future.
	\end{abstract}
	
	\section{Introduction}
	Satellite system has lots of advantages such as wide coverage, large communication capacity and low transmission latency \cite{k1} \cite{k2}. In response to the disadvantage of limited terrestrial network coverage, difficulty in supporting high-speed mobile user applications, and vulnerability to natural disasters \cite{belmekki2022unleashing}, satellite systems have good prospects, whether to supplement terrestrial communications or independently establish and cover the world. Therefore, within the next five years, a number of satellites will be launched into orbit, according to proposals presented by many satellite companies, such as SpaceX, Amazon, and OneWeb \cite{k3}.
	
    \par
	The following is the structure of this paper based on the description provided above. We start by going into further details about the benefits and drawbacks of LEO satellites. Then, we explain the motivation for applying SG to LEO satellite networks. In addition, a non-technical discussion of models of satellite location distribution, typical parameters, and key performance measures is provided in this paper, which reviews the analysis framework of some recent works based on SG. And most importantly, we derive some equations of T-S link in satellite-relayed communication system. Finally, a simulation setup is suggested with valuable system-level conclusions to demonstrate the potential benefits of employing SG in modeling and assessing LEO satellite communication systems.
	
	\subsection{Advantages and Challenges of LEO Satellite System}
	Satellites are classified into high earth orbit (geostationary earth orbit, GEO), medium earth orbit (MEO) and low earth orbit satellites (LEO) according to their orbital altitude \cite{k4}. With regard to capacity, coverage, transmission delay, system design, maintenance costs, and the challenge of frequency coordination, Table I explicitly examines the benefits and drawbacks of the high orbit satellite, low orbit satellite, and ground system.

\begin{table*}[]		
\centering
	\caption{Comparison of GEO-satellite, LEO-satellite and Ground System }
	\label{table_1}
	\begin{scriptsize}
	\begin{tabular}{|c|c|c|c|c|}
		\hline
		Type of System &
		Capacity &
		Coverage area &
		Costs &
		Latency \\ \hline
		LEO &
		Large &
		Comprehensive seamless coverage &
		\begin{tabular}[c]{@{}c@{}}High construction cost \\ and high maintenance cost\end{tabular} &
		Medium \\ \hline
		GEO &
		Limited &
		\begin{tabular}[c]{@{}c@{}}Large coverage of single satellite \\ and blind area of two poles\end{tabular} &
		\begin{tabular}[c]{@{}c@{}}Low construction cost \\ and low maintenance cost\end{tabular} &
		Large \\ \hline
		\begin{tabular}[c]{@{}c@{}}Terrestrial \\ Network\end{tabular} &
		Large &
		\begin{tabular}[c]{@{}c@{}}Small coverage of single base station and \\ having difficult in the communication of specific terrain\end{tabular} &
		\begin{tabular}[c]{@{}c@{}}Flexible construction \\ and very high maintenance costs \\ in remote areas\end{tabular} &
		Low \\ \hline
	\end{tabular}%
	\end{scriptsize}
\end{table*}

Despite a single GEO satellite that can provide a large coverage area, the conventional high orbit geostationary satellite system has blind spots at the poles. Compared to it, a constellation of LEO satellites can provide much lower latency and seamless coverage of many partially connected or unconnected areas of the globe, offering unparalleled advantages in terms of equality of access.

Although more LEO satellites are needed for seamless global connectivity, new space carriers have fortunately reduced launch costs and also assembly line production has reduced the cost of producing satellites.

The coverage of terrestrial systems is limited by the terrain. In remote areas where infrastructure is lacking, existing terrestrial networks have difficulty providing coverage because of the high cost of extending fiber to these areas. Therefore, LEO satellites offer better free-space transmission conditions and broader coverage compared to terrestrial networks. However, the design and performance analysis of the satellite constellation is particularly important because of the long deployment time and less flexibility of the satellites.

However, in the development of satellite systems, frequency resources are becoming the focus of competition among countries and satellite companies. According to current international rules, only one satellite in orbit within a constellation gets priority for frequency and orbital resources. By developing a satellite Internet, it can easily evolve into a competition for satellite frequency resources.

The Ku and Ka frequency bands, which are used by GEO satellite constellations, are also used by LEO satellite constellations. ITU regulations state that GEO satellites have priority when it comes to frequency use, but when LEO satellites travel through the area between a GEO satellite and its customers or gateways, there is an increased danger of LEO interference with the GSO satellite. In addition, there may be interference between LEO satellites and ground mobile communications, as well as interference amongst LEO constellation systems. The interference process is also particularly complicated since LEO satellites are in a movable condition.

	\subsection{Motivations of Applying SG in LEO Satellite Networks}
Stochastic geometry(SG) is a rich branch of applied probability theory, particularly suited to the research of random phenomena in the plane or in higher dimensions. SG has been widely used as a powerful mathematical tool in modeling and analysis of communication networks, because it is suitable for wireless networks’ inherent spatial randomness and uncertainty of signals \cite{k11}. Meanwhile, it is easier for researchers to use and understand for its closed-form expressions. And most importantly,  it has been proved that the lower bound of coverage probability and average achievable rate in the SG model is as tight as the upper bound of the regular mesh model \cite{k5}, which means SG has much higher efficiency. The advantages of SG-based modeling and analysis in LEO satellite systems are as follows.

First, it is clear from the above survey that the number of LEO satellites will explode in the future. For such a large system, fine-grained modeling of each LEO satellite requires a lot of effort. Performing a system-level analysis based on SG is much easier than defining and studying the behavior of each satellite. SG modeling does not need to rely on orbit metrics and the shape of a particular constellation. The explosive growth of dynamic networks requires a generalized and flexible stochastic network \cite{k6}.
	
	In addition, it is clear from the above analysis of the future challenges of the LEO satellite system that the research of the impact of interference on the system performance will be an important aspect, but other than SG there is still a lack of practical analysis tools to model the interference caused by satellites.
	
	Besides, SG is applicable to modeling and analysis of irregular topological networks \cite{k7}. In the existing non-SG models, satellites are often deployed in regular circular or hexagonal cells with the same coverage \cite{k8}. However, the distribution of satellites depends on latitude, and the coverage of satellites varies greatly in most cases. Therefore, the results of traditional analysis methods may differ significantly from the actual situation.
	
	And eventually, for the LEO satellite system, the performance of the SG model closely matches with that of the deterministic constellation in coverage probability and average achievable rate \cite{k9}.

	\section{SG-based Analytical Framework}

	\subsection{Models of Satellite Locations Distribution}
	In most of the available literature, satellites as well as users, satellite gateways(GWs) are assumed to be independent and uniformly distributed.
	
	For Poisson point process (PPP), the number of points in a predefined region is Poisson distributed, while their locations are uniformly distributed within the region. As one of the most commonly used point processes, PPP can fit the ground network well in performance analysis \cite{lou2021green}. And the coverage radius of GWs is negligible compared to the surface area of the earth, so GWs can be considered as points located in a large two-dimensional plane, PPP is suitable for modeling their locations \cite{k12}.
	
	Although PPP is highly practical, it is not the best choice for modeling finite area networks with a finite number of nodes because the satellite positions are on a sphere and the number of satellites is fixed for deterministic constellations. Therefore, modeling satellite positions as a binomial point process (BPP) is an effective solution \cite{k9}, \cite{wang2022evaluating}. Specifically, it places a fixed number of points on a sphere of fixed height. The zenith (elevation) and azimuth angles are uniformly distributed.
	
	In summary, the current main distribution model is that LEO satellites form independent BPP, distributed on different spheres, while terrestrial GWs form PPP.
	
	\subsection{Channel Model}
	The linear expression of the received power is $  \rho\eta C r^{-\alpha} $, where $\rho$ is used to describe the transmitted power and antenna gain, $\eta$ is used to describe large-scale fading, C is used to describe small-scale fading, and $ \alpha $ is called the path loss exponent.
	
	In the downlink transmission, the signal power transmission always uses this model: satellites have a larger gain in a particular direction realized by beamforming (BF) technology \cite{zhang2022doa}. The GW or user can receive main-lobe power in this model, while the power from the interfering source tends to be side-lobe, which can improve the signal-to-interference ratio (SIR) of the system \cite{k13}.
	
	The path loss exponent is a number between 2 and 4 usually. 
	
	Large-scale fading is usually modeled as a log-normal shadowing to describe additional gain. Some of the literature takes the attenuation of air absorption caused by the resonance of water vapor into consideration \cite{k12}\cite{k13}, which is called rain attenuation. It changes with geographical location, such as desert and rain forest.
	
	Small-scale fading is a random variable that denotes the channel fading power gains. In ideal free space propagation, small-scale fading does not occur. Shadowed-Rician (SR) fading is the most accurate channel fading model and is widely applied to make accurate performance estimates \cite{k14}\cite{k15}, especially for space-to-ground links \cite{k12}.
	
	When LEO satellites rely on a ground GW as a relay, a user is covered by a satellite when both satellite-GW and GW-user links achieve the SINR requirements \cite{k12}.
	
	\subsection{Contact Distance}
	The distance between the user or GW and the connected (service-providing) satellite is known as the contact distance. Two expressions of the contact distance distribution for BPP are given in reference \cite{k9}. The contact distance is the distance between the user and the closest satellite at a specific height, calculated using the strongest average received power association approach. As a random variable, the complementary cumulative distribution function (CCDF) of contact distance at $do$ is equal to the probability that there are no satellites in the spherical cap at the intersection of spherical surface over which satellites are distributed and the cone with side length $do$ centered at the location of the user/GW region closer to the user. 
	
	\subsection{Coverage Probability}
	In this section, we study the application of SG in system performance analysis. Two related definitions for measuring system performance are:
    \begin{itemize}
    	\item The  signal-to-interference-plus-noise ratio (SINR)  is used to measure the communication quality of the system.
    	\item Coverage probability is defined as the prob-ability that the SINR is larger than a prede-termined acceptable threshold. It represents the probability that the system can provide reliable connections.
    \end{itemize}
    Once the contact distance has calculated the distance to the connected satellite, the interference is the total power of all other satellites above the horizon. The coverage probability can then be computed after obtaining the SINR.
    Different application contexts use different expressions to describe coverage probabilities. A user is covered by a satellite when both the satellite-GW and GW-user links meet the SINR requirements when LEO satellites use a ground GW as a relay \cite{k12}.

	\section{Original Work}
	
	\subsection{System Model}
    \begin{figure}[H]
	\centering
	\includegraphics[width=0.7\linewidth]{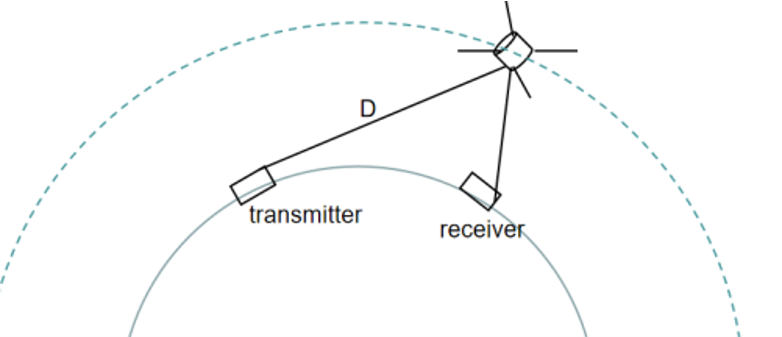}
	\caption{T-S-R Model}
	\label{fig:tsr}
    \end{figure}
    
	In the past, many research models studied the communication from satellite to gateway and then to user, namely downlink communication. As for our group, we firstly use spherical stochastic geometry in satellite-relayed communication model, which had never been used before. That is, instead of assuming the gateway as a transmission point of information \cite{k12}, a model such as transmitter-satellite-receiver(T-S-R) (Figure.\ref{fig:tsr}) can be analyzed independently and individually. Based on SG, we assume that the relayed satellite satisfies the BPP distribution on the low earth orbit, and the transmitter as well as the receiver on the ground satisfy the PPP distribution. This forms the SG-based analytical framework of our group. And our derivation mainly focus on finding the distance between the best relay satellite and the transmitter(D) \cite{k16}. The specific derivation steps maily refer to the steps in \cite{k12}.
	\par
	For our model is a simplest T-S-R relay communication model based on SG in coverage probability calculation. Such set is independent from each other, that is, we can easily extend this set of methods to multi-satellite communication chain, which is shown in the following simulation section(T-S-S-R model), and even multi-GW multi-S link communication quality analysis in the future.
	
\subsection{The Obtained Formula}
	All assumptions as well as derivations are made in order to eventually be able to calculate the coverage probability: 
	\begin{equation}
		P_{\text {cov }}=P_{\text {cov }}^{T-Sat} P_{\text {cov }}^{Sat-Re},
	\end{equation}
    where 
    \begin{equation}
     P_{\mathrm{cov}}^{T-Sat}=P\left(\frac{\rho_{r}^{t}}{\sigma_{s}^{2}} \geq \gamma_{s}\right),
    \end{equation}
    
    \begin{equation}
     P_{\mathrm{cov}}^{Sat-Re}=P\left(\frac{\rho_{r}^{s}}{\sigma_{r}^{2}} \geq \gamma_{r}\right),
    \end{equation}
    ${\sigma }_{s}^{2}$  and ${\sigma }_{r}^{2}$ are the noise powers at the satellite and the receiver, respectively \cite{k12}.

To simplify the model, we assume that the communication from the transmitter to the satellite is influenced by the same factors as the communication from the satellite to the receiver, and the other parameters should be the same excluding the influence of such a factor as distance. Therefore, first of all, we try to deduce the contact distance. Here we only discuss the distance from the transmitter to the satellite (D), for the distance from the satellite to the receiver can be deduced in the same way. The globe uses a spherical coordinate system, so we first use angles to express geometric relations, and then use the law of cosines to convert angles into distances.

\begin{figure}[H]
	\centering
	\includegraphics[width=0.7\linewidth]{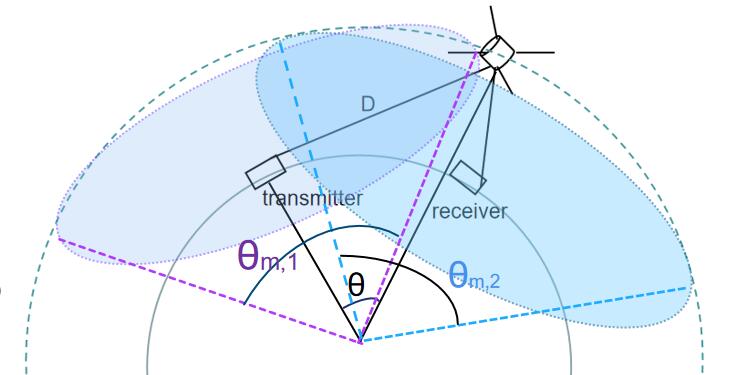}
	\caption{Satellite-relayed communication model (T-S-R)}
	\label{fig:theta}
\end{figure}

	As shown in Figure.\ref{fig:theta},  we suppose $  \theta  $ is the angle between the transmitter and the satellite and the center of the earth. The derivation is as follows:
	\begin{itemize}
		\item  Fix the receiver's position,  its dome angle ${\theta }_{m,2}$ and coverage area ${A}_{2}$
		\item 	Change the position of the transmitter, its dome angle ${\theta }_{m,1}$ and coverage area ${A}_{1}$ will change too \cite{wang2022stochastic}. Under such changes, there is an overlap between ${A}_{1}$ and ${A}_{2}$.
		\item	According to the distribution of satellites, judge whether the transmitter can find a satellite that can successfully communicate in the overlapping area and calculate the probability of no satellite.
		\item	The coverage probability is the probability that N-1 satellites are not in the overlap area. Or we can call it cumulative distribution function (CDF) of contact angle. Then by using the law of cosines we replace theta with D and take the derivative of D, we can get the probability density function(PDF) of contact distance distribution.
	\end{itemize}
\begin{equation}
	\begin{split}
		&f=\frac{1}{\sigma_{2}} 
		N\left(1-\frac{\int_{\sigma_{5}}^{\sigma_{1}} \sigma_{6} \mathrm{~d} l+\int_{\sigma_{8}}^{R \sin \left(\sigma_{17}\right)} \sigma_{6} \mathrm{~d} l}{\sigma_{2}}\right)^{N-1}\\
		&\times \Bigg(   \int_{\sigma_{5}}^{\sigma_{1}} \sigma_{4} \mathrm{~d} l+\int_{\sigma_{8}}^{R \sin \left(\sigma_{17}\right)} \sigma_{4} \mathrm{~d} l+2 R \operatorname{asin}\left(\frac{\sqrt{\sigma_{13}-R^{2} \sigma_{11}^{2} \cos \left(\sigma_{17}\right)^{2}}}{R}\right)\\
		&\times \left(\frac{\mathrm{D} \sigma_{11} \sin \left(\sigma_{17}\right)}{R_{e}}+R \cos \left(\sigma_{17}\right)\left(\frac{\frac{\sigma_{3}}{2 \sigma_{15}}-\frac{2 \mathrm{D} \sigma_{16} \sin \left(\sigma_{17}\right)}{R R_{e}}}{\cos \left(\sigma_{17}\right)}+\frac{\mathrm{D} \sin \left(\sigma_{17}\right) \sigma_{14}}{\sigma_{7}}\right)\left(\sigma_{11}^{2}+1\right)\right)\\
		& -2 R^{2} \cos \left(\frac{\theta_{m, 2}}{2}\right) \operatorname{asin}\left(\frac{\sqrt{\sigma_{13}-R^{2} \sigma_{9}^{2} \sigma_{18}}}{R}\right)\left(\sigma_{9}^{2}+1\right)\\
		&\times \left(\frac{\sigma_{3}}{2 \cos \left(\sigma_{17}\right) \sigma_{15}}+\frac{\mathrm{D} \sin \left(\sigma_{17}\right) \sigma_{12}}{\sigma_{7}}\right) -\frac{2 \mathrm{D} R \operatorname{asin}(0) \cos \left(\sigma_{17}\right)}{R_{e}}\Bigg)
	\end{split}
\end{equation}
where $\sigma_{1}-\sigma_{18}$ are shown in Appendix A.

In addition to distance, there are other attenuation factors, such as rain attenuation, which we won't discuss too much in this section. We end up with the expression:
    \begin{equation}
    	\begin{aligned}
    		P_{\mathrm{cov}}^{\mathrm{S}-\mathrm{Sat}}=& \int_{0}^{\infty} \frac{1}{2 \sqrt[4]{D}} f_{D 1}(\sqrt[4]{D}) \\
    		&-\left(\frac{2 b_{0} m}{2 b_{0} m+\Omega}\right)^{m} \sum_{z=0}^{\infty} \frac{(m)_{z}}{z ! \Gamma(z+1)}\left(\frac{\Omega}{2 b_{0} m+\Omega}\right)^{z} \\
    		& \times \int_{0}^{\infty} \gamma\left(z+1, \frac{1}{2 b_{0}} c \sqrt{D}\right) \frac{1}{2 \sqrt[4]{D}} f_{D 1}(\sqrt[4]{D}) d D
    	\end{aligned}
    \end{equation}
    \begin{equation}
    	\begin{aligned}
    		P_{\mathrm{cov}}^{\mathrm{Sat}-\mathrm{Re}}=& \int_{0}^{\infty} \frac{1}{2 \sqrt[4]{D}} f_{D 1}(\sqrt[4]{D}) \\
    		&-\left(\frac{2 b_{0} m}{2 b_{0} m+\Omega}\right)^{m} \sum_{z=0}^{\infty} \frac{(m)_{z}}{z ! \Gamma(z+1)}\left(\frac{\Omega}{2 b_{0} m+\Omega}\right)^{z} \\
    		& \times \int_{0}^{\infty} \gamma\left(z+1, \frac{1}{2 b_{0}} c \sqrt{D}\right) \frac{1}{2 \sqrt[4]{D}} f_{D 1}(\sqrt[4]{D}) d D
    	\end{aligned}
    \end{equation}
where $ \Gamma(\cdot) $ denotes the gamma function, $ \gamma(\cdot, \cdot) $ is the lower incomplete gamma function, $ (m)_{z} $ is the Pochhammer symbol, while $m$ , ${b}_{0}$ and $\Omega $ are the parameters of the SR fading \cite{k12}.

	\section{SG-based Simulation Setup and Numerical Results}
	
	\subsection{Simulation Model}
	
	We simulate this stochastic geometry process with the Monte Carlo method to evaluate the coverage probability of satellite-relayed system.
	
	Assume that there are gateways conforming to a PPP with a surface density of ${\lambda }_{GW}$ distributed over the Earth's surface. Assume that a fixed quantity of LEO satellites conforming to a BPP distribution are distributed in a fixed LEO. Consider a gateway-satellite-gateway(T-S-R) downlink relay communication system with the same receiving antenna and transmitting antenna of the gateway. And consider that there is an overlap between the region swept by the width of the wave flaps of the receiving and transmitting station antennas to establish a link. Rain fading, propagation fading, and antenna gain for signal strength occur in both T-S and S-R links. In the S-S link only propagation attenuation and antenna gain occur. And it is assumed that the antenna gain, the Earth's surface, and the background noise of the space in which the LEO satellite is located are a constant.
\begin{table*}
\centering
	\caption{System Parameters}
	\label{table_2}
		\begin{tabular}{|c|c|}
			\hline
			\textbf{PRAMETER}                                        & \textbf{VALUE}      \\ \hline
			Surface background noise                                 & -80dB               \\ \hline
			Background noise in space 550km from the earth's surface & -100dB              \\ \hline
			Average rain attenuation                                 & -2dB                \\ \hline
			Lobe angel of the ground-based antenna                   & $\frac{65}{180}\pi $                 \\ \hline
			Ground gateway antenna gain                              & 80dB                \\ \hline
			Satellite antenna gain                                   & 60dB                \\ \hline
			SR fading($\Omega $, ${b}_{0}$, $m$)                                        & SR(1.29, 0.158, 19.4) \\ \hline
			Carrier frequency                                        & 300MHz              \\ \hline
		\end{tabular}%
\end{table*}
	
	\subsection{Simulation Details}
	\begin{enumerate}
		\item \emph{Selection of relay satellite: }To ensure the successful transmission of the transmit signal to the relay satellite as much as possible, the relay satellite is selected to be the closest satellite to the transmitting station in the effective overlap region of the wave flaps between the transmitting and receiving stations.
		\item \emph{SR fading: }the following lemma from \cite{k12}:
		\begin{equation}
			F_{D}(d) = P(D<d)=1-\prod_{i=1}^{n} P\left(D_{i} \geq d\right),
		\end{equation}
	    Since it contains the summation using the 50th order incomplete gamma function, the inverse function cannot be taken directly. So it is first fitted with a 10th order polynomial in the function value (0-1) increasing curve (rounding off both ends), after which the inverse is taken, after which the SR distribution can be generated using uniform random numbers
	    
		\item \emph{Simulation of spherical BPP and PPP: }The total number of random points of PPP is obtained from the Poisson distribution in one dimension, after which the random points still conform to BPP, and the following equation is obtained by taking the inverse function of the CDF of the sphere coordinates of the uniformly distributed points using the sphere. Then use the uniform random number to generate the point process.
		
		\item \emph{Calculation of coverage probability: }The probability that the ratio of transmitter's transmit power to receiver's receive power is greater than the threshold gamma is the coverage probability.
		In this paper, the coverage probabilities of T-S-R process and S-R process with valid link occurrence, and the coverage probabilities of T-S-R process and T-S-S-R process with arbitrary gateway communication without ensuring valid link occurrence are considered respectively.
		
		\item \emph{The practical optimization of the introduction of satellite-satellite links for the premise of not ensuring the effective occurrence of the link: }
		\begin{figure}[H]
			\centering
			\includegraphics[width=0.7\linewidth]{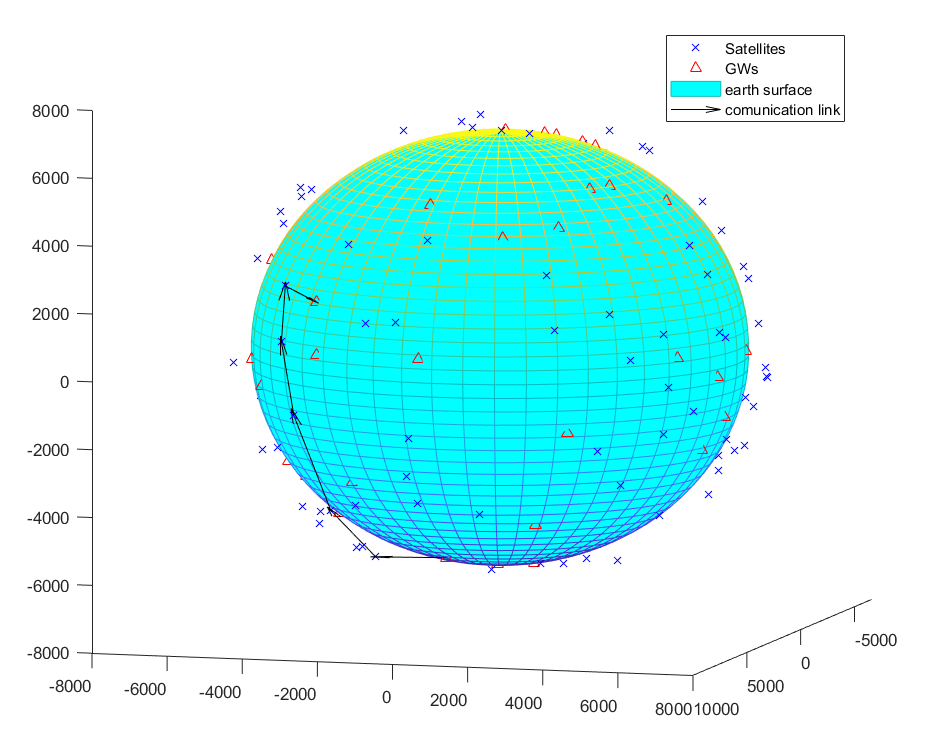}
			\caption{Satellite communication(T-S-S-R model)}
			\label{fig:tssr}
		\end{figure}

		The introduction of a satellite-satellite communication link makes it possible for any two gateways on the earth's surface to establish a communication link. The signal is sent from the transmitting station to the nearest satellite, through the satellite-satellite link, and finally to the receiving station when it reaches within the antenna flap width of the target receiving station. Although there is some fading and delay, this greatly increases the coverage probability of establishing communication between any gateways. Where the satellite link selection is obtained by a local greedy algorithm.
		The spherical coordinate system is transformed into the two-dimensional matrix shown in the figure. Set the local distance threshold ${d}_{max}$, the current starting point to the end point vector $\nu -t\mathrm{arg}et$, and find the satellite point location within the range that makes the smallest angle with $\nu -t\mathrm{arg}et$ as the new starting point each time, and iterate to get the satellite link.
		
	\end{enumerate}

    \subsection{Comparison and Analysis of Numerical Results}
    \begin{figure}[H]
    	\centering
    	\includegraphics[width=0.7\linewidth]{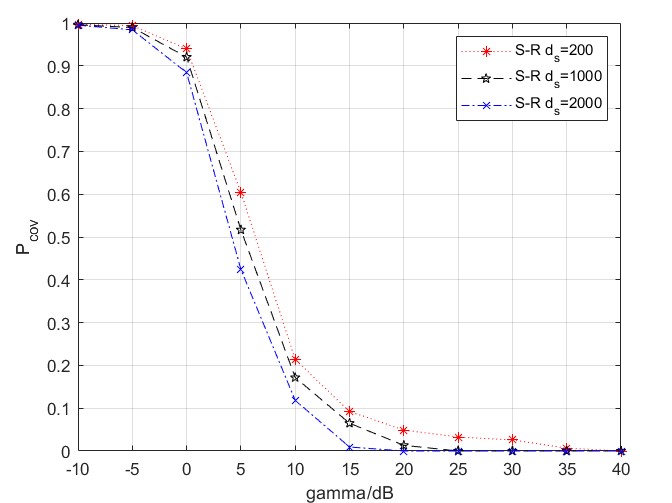}
    	\caption{${\lambda }_{GW}=1.96{e}^{-7}/k{m}^{-2}$; ${\theta }_{m}=\frac{65}{180}\pi $; ${N}_{S}=100000$}
    	\label{fig:q1}
    \end{figure}
    \begin{figure}[H]
    	\centering
    	\includegraphics[width=0.7\linewidth]{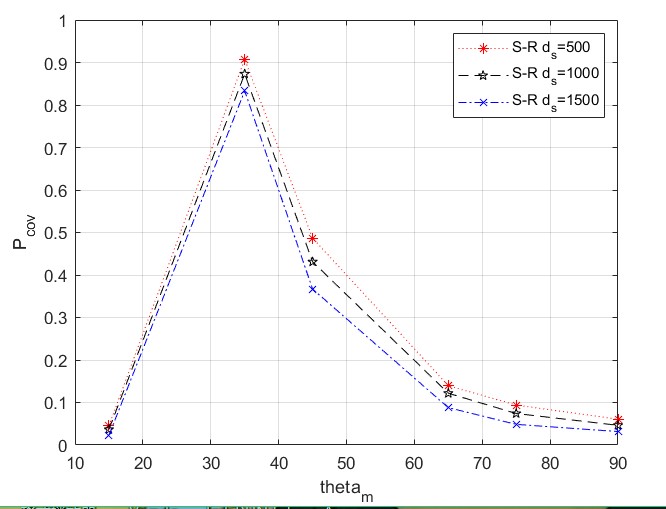}
    	\caption{$\gamma =12dB$; ${\lambda }_{GW}=1.96{e}^{-7}/k{m}^{-2}$}
    	\label{fig:q2}
    \end{figure}
    \begin{figure}[H]
    	\centering
    	\includegraphics[width=0.7\linewidth]{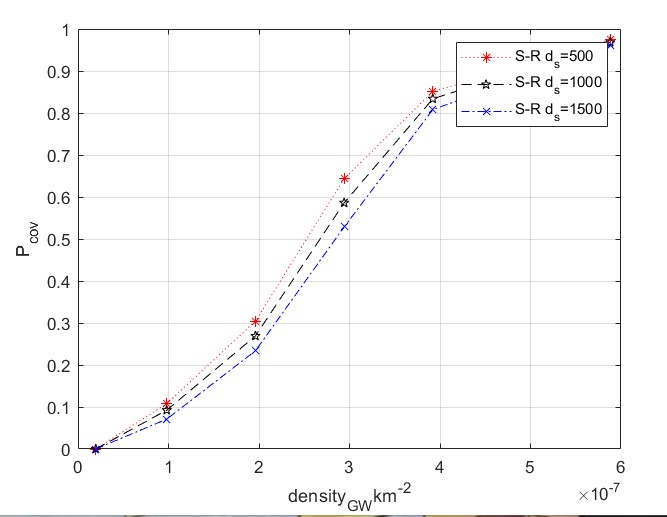}
    	\caption{$\gamma =12dB$;${N}_{S}=100000$}
    	\label{fig:q3}
    \end{figure}
    From the Figure \ref{fig:q1} \ref{fig:q2} \ref{fig:q3} it can be seen that the effect of changing ds alone on the S-R coverage probability of a single relay communication is not significant when Ns is fixed to a certain value. From the subsequent analysis we can learn that ds is strongly correlated with Ns.
    
    \begin{figure}[H]
    	\centering
    	\includegraphics[width=0.7\linewidth]{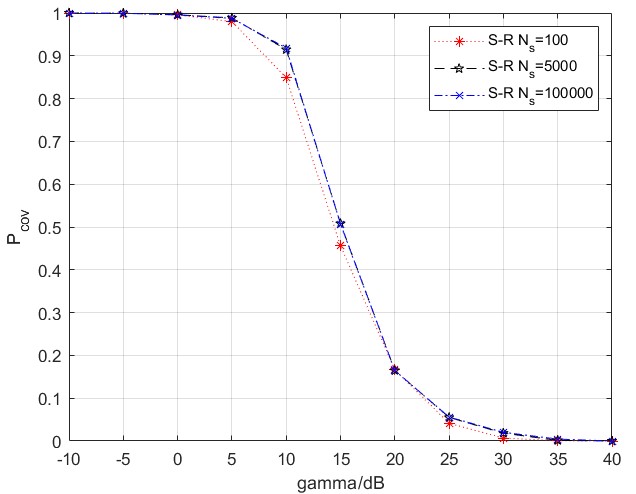}
    	\caption{$ds=550km$; ${\lambda }_{GW}=1.96{e}^{-7}/k{m}^{-2}$}
    	\label{fig:q4}
    \end{figure}
    \begin{figure}[H]
    	\centering
    	\includegraphics[width=0.7\linewidth]{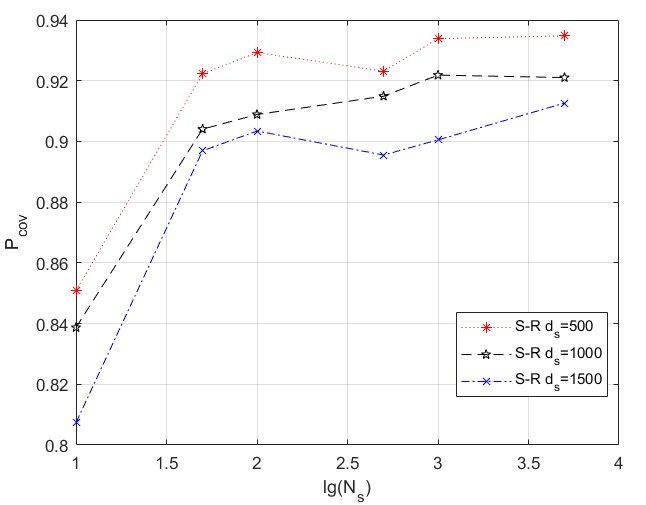}
    	\caption{$\gamma =12dB$; ${\lambda }_{GW}=1.96{e}^{-7}/k{m}^{-2}$}
    	\label{fig:q5}
    \end{figure}
    Figure \ref{fig:q4} illustrates that with ds=550km; densityGW=100/4piRe2, the number of satellites is saturated for a single communication to be able to achieve complete coverage. However, it should be noted that the saturation of coverage probability only indicates that the communication quality must be guaranteed, but it does not guarantee the time delay. Because once the regional satellites are fully occupied, the waiting time will be huge.
	
	Figure \ref{fig:q5} shows that the number of satellites in fixed orbits, the lower the orbital altitude the higher the coverage probability. The coverage probability increases with a fixed orbital altitude and a higher number of orbiting satellites, but saturates after reaching a certain value, the same result as shown in Figure \ref{fig:q1} \ref{fig:q2} \ref{fig:q3}. It is also clearly shown that as the orbital altitude increases, the number of satellites required to reach the same coverage probability increases rapidly.
	
	\begin{figure}[H]
		\centering
		\includegraphics[width=0.7\linewidth]{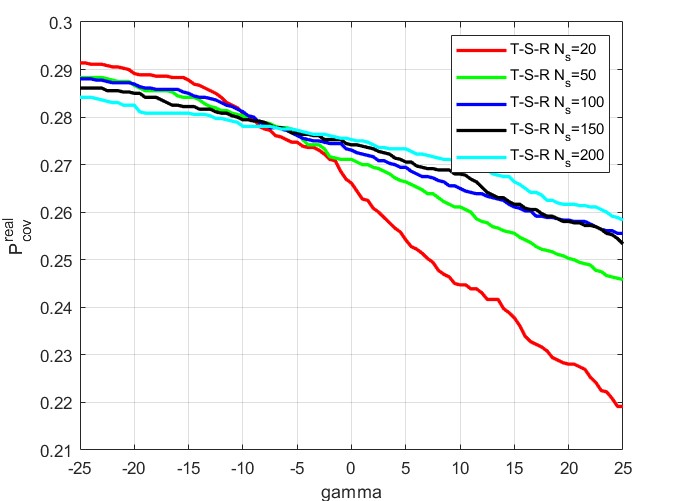}
		\caption{T-S-R coverage probability between arbitrary gateways}
		\label{fig:q6}
	\end{figure}
	\begin{figure}[H]
		\centering
		\includegraphics[width=0.7\linewidth]{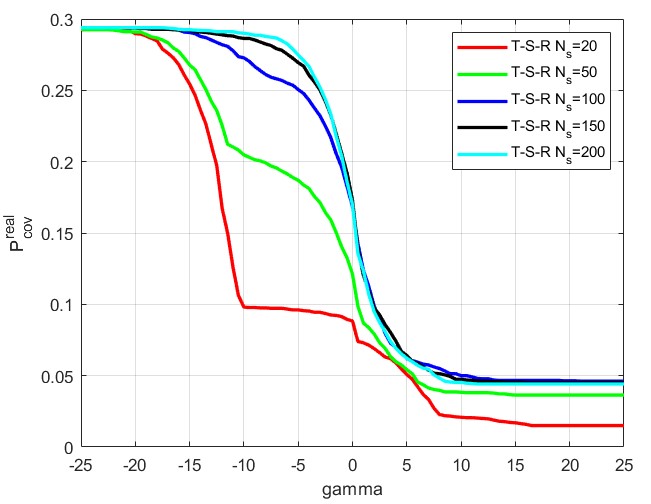}
		\caption{T-S-S-R coverage probability between arbitrary gateways}
		\label{fig:q7}
	\end{figure}
	\begin{figure}[H]
		\centering
		\includegraphics[width=0.7\linewidth]{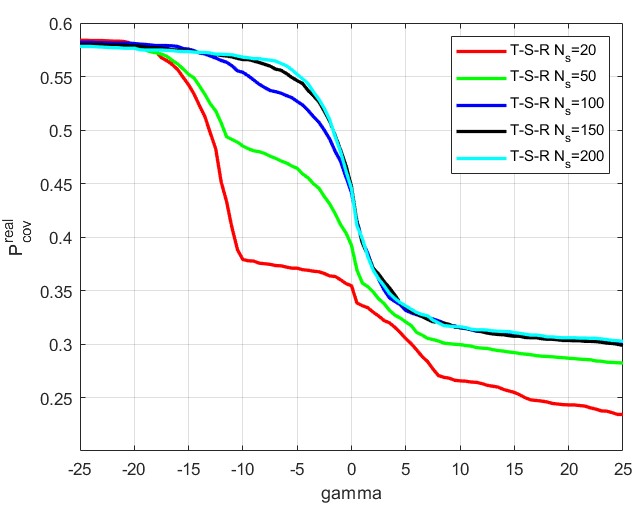}
		\caption{Coverage probability between arbitrary GWs(T-S-R+T-S-S-R)}
		\label{fig:q8}
	\end{figure}
	The previously obtained results are coverage probabilities calculated from the SNR of the communication process under the premise of ensuring the overlap between the receiving station and the transmitting station's flap width sweep area, while here the coverage probability between any two gateways is analyzed. It can be clearly seen that the contribution of the communication link between the satellites to the coverage probability is significant, and this contribution increases with the number of satellites and saturates after reaching a certain value.
	
	\begin{figure}[H]
		\centering
		\includegraphics[width=0.7\linewidth]{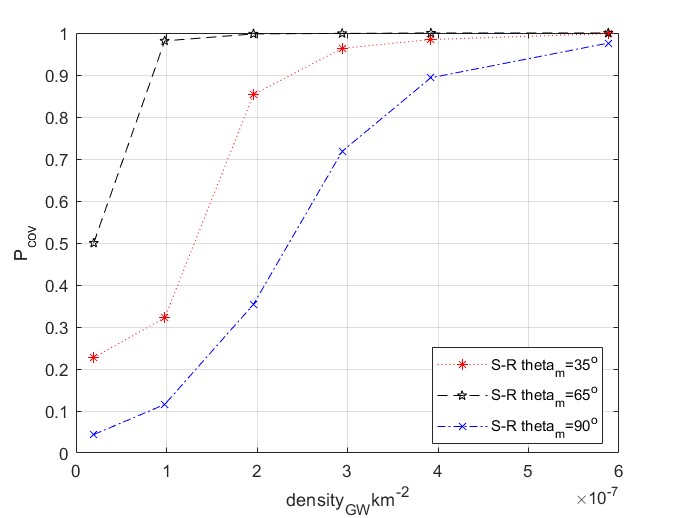}
		\caption{$\gamma =12dB$; $ ds=550km $}
		\label{fig:q9}
	\end{figure}
	\begin{figure}[H]
		\centering
		\includegraphics[width=0.7\linewidth]{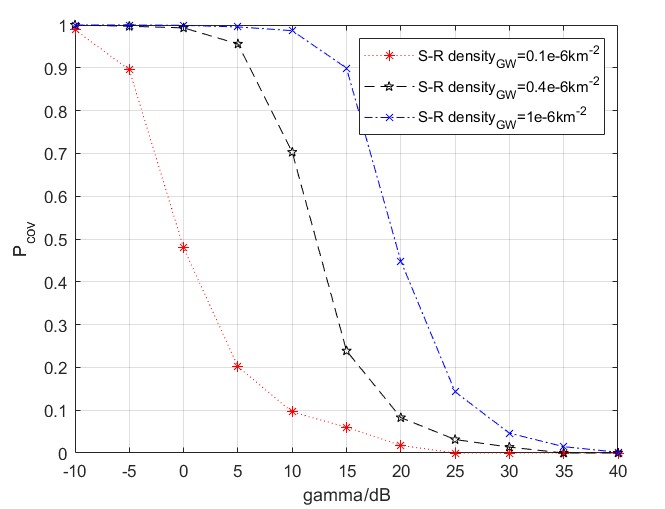}
		\caption{${\theta }_{m}=\frac{65}{180}\pi $; $ Ns=100000 $; $ ds=550km $}
		\label{fig:q100}
	\end{figure}
	Figure \ref{fig:q9} shows that the width of the lope required to achieve the same coverage probability increases as the ground gateway density increases. and that the coverage probability increases significantly as the lope decreases, and that the coverage probability increases significantly as the gateway density increases.
	
	Figure \ref{fig:q100} shows more clearly in case of ${\theta }_{m}=\frac{65}{180}\pi $; $ Ns=100000 $; $ ds=550km $, the relationship between coverage probability, threshold and density.
	
	\begin{figure}[H]
		\centering
		\includegraphics[width=0.7\linewidth]{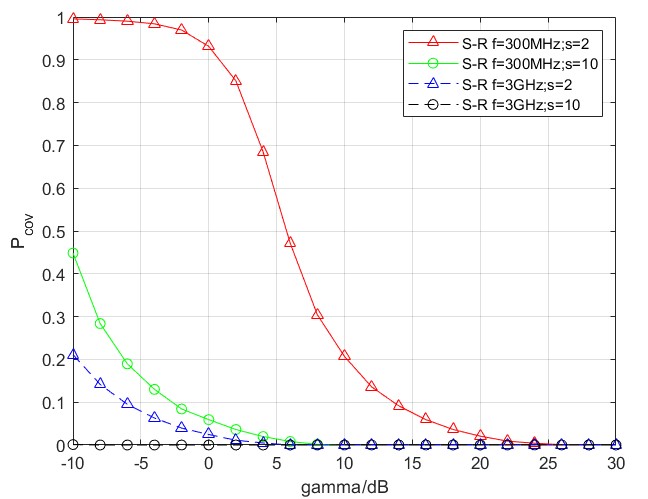}
		\caption{${N}_{GW}=100$; $ Ns=100 $}
		\label{fig:q10}
	\end{figure}
	Figure \ref{fig:q10} shows the effect of carrier frequency and rain attenuation on the coverage probability.
	The rain attenuation coefficient at this point depends only on the depolarization process of raindrops for the carrier, because the scattering effect of raindrops can be neglected due to the long wavelength of the UHF band we have chosen. We can obviously under the existing model assumptions, the increase of frequency and the enhancement of rain attenuation effect is significant for the coverage probability. However, in the actual process, the effect brought by the change of frequency can be improved due to the enhancement of the gain circuit for the amplification effect of high frequency.
	
\begin{table*}[]
\centering
	\caption{Table of Notations and Acronyms}
	\label{table_3}
	\begin{small}
	\begin{tabular}{c|c}
	    \hline
		\textbf{Notations}   & \textbf{Description}                        \\ \hline
		LEO, GEO             & Low, Geostationary Earth Orbit                \\ \hline
	    SG  & Stochastic Geometry  \\ \hline
		GW    & Gateway           \\ \hline
		BPP, PPP   & Binomial Point Process, Poisson Point Process \\ \hline
		SINR      & Signal-to-interference-plus-noise ratio \\ \hline
		SIR      & Signal-to-interference ratio   \\ \hline
		SR fading   & Shadowed-Rician fading  \\ \hline
		$P_{\mathrm{cov}}^{T-Sat}$, $P_{\mathrm{cov}}^{Sat-Re}$ &
		Coverage probability for T-S link, S-R link \\ \hline
		$\rho_{r}^{t}$, $\rho_{r}^{s}$ &
		Transmitter power, Satellite power \\ \hline
		$\sigma_{s}^{2}$, $\sigma_{r}^{2}$ &
		Noise power at the satellite, at the receiver \\ \hline
		$\gamma_{s}$, $\gamma_{r}$ &
		SNR threshold at the satellite, at the receiver \\ \hline
		D  & Length of T-S link     \\ \hline
		R, Re    & The radius of the sphere, the Earth     \\ \hline
		${\theta }_{m}$      & The Dome angle    \\ \hline
		${A}_{1}$, ${A}_{2}$ & Shaded area of transmitter, receiver     \\ \hline
		$  \theta  $ &
		\begin{tabular}[c]{@{}c@{}}The angle between the transmitter and the satellite and the center of the earth\end{tabular} \\ \hline
		CDF       & Cumulative distribution function      \\ \hline
		PDF    & Probability density function   \\ \hline
		${\lambda }_{GW}$    & The surface density of GW in PPP    \\ \hline
		$ds$    & Altitude of LEO   \\ \hline
		${N}_{S}$    & Number of satellites in fixed orbit     \\ \hline
		$\gamma$     & Equals to $\gamma_{s}$ or $\gamma_{r}$     \\ \hline
		f  & Carrier frequency   \\ \hline
		s  & Average rain attenuation  \\ \hline
		\end{tabular}
	\end{small}
 \end{table*}
	
\section{Conclusions and Future Applications}
This article investigates the application of SG to simulating and analyzing LEO satellite communications systems, introduces several satellite distributions, and analyzes the channel models in the literature. From a non-technical perspective, we describe contact distance and coverage probability. Then, we use the satellite-relayed communication model based on SG to derive the PDF of the contact distance between the transmitter and the satellite(T-S link) and the probability of successful communication (coverage probability) in T-S-R model. A series of simulations are made on this model and the following conclusions are obtained. Increasing the number of satellites and the height of the constellation within a specific range can effectively promote the coverage probability. The ability of the satellites to communicate with each other has a considerable effect on promoting the coverage probability. This indicates that the recent technological advances in enabling S-S communications will significantly widen the set of applications that can rely on LEO satellite communications.

High altitude platforms (HAPs) are airborne base stations placed in the stratosphere by airships or drones, combining the advantages of satellite communication systems and ground mobile cellular networks. Due to the greater distance from the ground, the number of satellites required to complete the communication coverage is less, and the coverage radius is larger than that of the ground mobile system, which can save costs and also accomplish a certain quality of high-capacity communication \cite{k10}. However, its disadvantages are also obvious: stronger rain fading attenuation, the need to use higher frequency bands; the multipath effect will affect the user's received power, and the path loss is large \cite{zhang2022doa2}. There is no relevant paper studying HAPs by using SG, but there is a high possibility of future research in this area.

\appendices
\section{The expressions of $\sigma$}
\begin{equation}
     \sigma_{1}=R \sin \frac{\theta_{m,2}}{2}
\end{equation}
    
\begin{equation}
    \sigma_{2}=4 \pi R^{2}
    \end{equation}

\begin{small}
\begin{equation}
	\sigma_{3}=\sqrt{2}\left(\frac{2 \mathrm{D} \cos \left(\sigma_{17}\right) \sin \left(\sigma_{17}\right)}{R R_{e}}-\frac{2 \mathrm{D} \cos \left(\frac{\theta_{m, 2}}{2}\right) \sin \left(\sigma_{17}\right)}{R R_{e}}+\frac{2 \mathrm{D} \cos \left(\frac{\theta_{m, 2}}{2}\right) \sigma_{16}^{2} \sin \left(\sigma_{17}\right)}{R R_{e}}\right)
\end{equation}
\end{small}

    \begin{equation}
	\sigma_{4}=-\frac{2 \mathrm{D} R \cos \left(\sigma_{17}\right) \sin \left(\sigma_{17}\right)}{R_{e} \sqrt{1-\frac{\sigma_{13}-l^{2}}{R^{2}}} \sigma_{10}}
\end{equation}
    \begin{equation}
	\sigma_{5}=-R \sigma_{9} \cos \left(\frac{\theta_{m, 2}}{2}\right)
\end{equation}
    \begin{equation}
	\sigma_{6}=2 R \operatorname{asin}\left(\frac{\sigma_{10}}{R}\right)
\end{equation}
    \begin{equation}
	\sigma_{7}=R R_{e} \cos \left(\sigma_{17}\right)^{2}
\end{equation}
    \begin{equation}
	\sigma_{8}=-R \sigma_{11} \cos \left(\sigma_{17}\right)
\end{equation}
    \begin{equation}
	\sigma_{9}=\tan \left(\cos \left(\frac{\theta_{m, 2}}{2}\right)+\frac{\sigma_{12}}{\cos \left(\sigma_{17}\right)}\right)
\end{equation}
    \begin{equation}
	\sigma_{10}=\sqrt{\sigma_{13}-l^{2}}
\end{equation}
    \begin{equation}
	\sigma_{11}=\tan \left(\cos \left(\frac{\theta_{m, 2}}{2}\right)-\frac{\sigma_{14}}{\cos \left(\sigma_{17}\right)}\right)
\end{equation}
    \begin{equation}
	\sigma_{12}=2 \cos \left(\frac{\theta_{m, 2}}{2}\right) \sigma_{16}-\sqrt{2} \sigma_{15}
\end{equation}
    \begin{equation}
	\sigma_{13}=R^{2} \sin \left(\sigma_{17}\right)^{2}
\end{equation}
    \begin{equation}
	\sigma_{14}=2 \sigma_{16} \cos \left(\sigma_{17}\right)-\sqrt{2} \sigma_{15}
\end{equation}
    \begin{equation}
	\sigma_{15}=\sqrt{\sigma_{18}+2 \cos \left(\frac{\theta_{m, 2}}{2}\right) \sigma_{16}{ }^{2} \cos \left(\sigma_{17}\right)-2 \cos \left(\frac{\theta_{m, 2}}{2}\right) \cos \left(\sigma_{17}\right)+\cos \left(\sigma_{17}\right)^{2}}
\end{equation}
    \begin{equation}
	\sigma_{16}=\operatorname{asin}\left(\frac{d}{2 R_{e}}\right)
\end{equation}
    \begin{equation}
	\sigma_{17}=\frac{-\mathrm{D}^{2}+R^{2}+R_{e}^{2}}{2 R R_{e}}
\end{equation}
    \begin{equation}
	\sigma_{18}=\cos \left(\frac{\theta_{m, 2}}{2}\right)^{2}
\end{equation}

\bibliographystyle{IEEEtran}
\bibliography{reference}

\end{document}